\input{aipcheck}

\documentclass[
    ,final            
  ]
  {aipproc}

\layoutstyle{8x11double}

\begin{document}

\title{Euclidean quantum gravity and stochastic approach: Physical reality of complex-valued instantons\footnote{A proceeding for the Multiverse and Fundamental Cosmology Conference (Multicosmofun'12). Talk on the 10th of September, 2012, Szczecin, Poland.}}


\author{Dong-han Yeom}{
  address={Center for Quantum Spacetime, Sogang University, Seoul 121-742, Republic of Korea}
}

\begin{abstract}
In this talk, we compare two states: the stationary state in stochastic inflation and the ground state wave function of the universe. We already know that, for the potential with a static field, two pictures give the same probability distribution. Here, we go beyond this limit and assert that two pictures indeed have deeper relations. We illustrate a simple example so that there is a corresponding instanton if a certain field value has a non-zero probability in the statistical side. This instanton should be complex-valued. Furthermore, the compact manifold in the Euclidean side can be interpreted as a coarse-graining grid size in the stochastic universe. Finally, we summarize the recent status and possible applications.
\end{abstract}

\maketitle

\section{Introduction}

In black hole physics, there are mainly two ways to calculate the Hawking temperature. One is to use the quantum field theory in the curved space-time background, while the other is to use the Euclidean analytic continuation. Similar to this, in cosmology, we also have two ways to include quantum fluctuations: one is to use the quantum field theory in the de Sitter background and this method is extended to the stochastic approach, while the other is the Euclidean quantum cosmology by using the mini-superspace and instantons.

In this presentation, we aim to compare two approaches for a simple case. Here, we use a simple Einstein gravity with a scalar field $\phi$
\begin{eqnarray}
S = \int d^{4}x \sqrt{-g} \left[ \frac{1}{16\pi}R - \frac{1}{2} (\nabla \phi)^{2} - V(\phi) \right],
\end{eqnarray}
where we consider a quadratic potential,
\begin{equation}
V(\phi) = V_{0} + \frac{1}{2} m^{2} \phi^{2} = V_{0} \left( 1 + \frac{1}{2} \mu^{2} \phi^{2} \right)
\end{equation}
with $\mu\phi \ll 1$ limit. We define the Hubble parameter $H = \sqrt{8\pi V_{0}/3}$. We can choose $V_{0}=1$ without loss of generality up to the parameter redefinition. We summarize the discussion of the author's recent paper \cite{Hwang:2012mf}, comment their physical meanings, and discuss possible future applications.

\section{Fuzzy instantons}

First, we briefly summarize the recent development in Euclidean quantum cosmology. In this area, people find a solution of the Wheeler-DeWitt equation. Especially, it is known that the Euclidean path integral can be regarded as the ground state of the wave function of the universe:
\begin{eqnarray}
\Psi[h_{\mu\nu}, \chi] = \int_{h, \chi} \mathcal{D}g\mathcal{D}\phi \;\; e^{-S_{\mathrm{E}}[g,\phi]},
\end{eqnarray}
where $h_{\mu\nu}=\partial g_{\mu\nu}$ and $\chi = \partial \phi$ are defined on a certain boundary of a compact Euclidean manifold. Here, we consider the Euclidean action
\begin{eqnarray}
S_{\mathrm{E}} = -\int d^{4}x \sqrt{+g} \left( \frac{1}{16\pi}R - \frac{1}{2} (\nabla \phi)^{2} - V(\phi) \right).
\end{eqnarray}
This path-integral is to sum over all non-singular geometries with a single boundary.

For the cosmological purposes, we assume the following metric:
\begin{eqnarray}
ds_{\mathrm{E}}^{2} = d\tau^{2} + \rho^{2}\left(\tau\right) d\Omega_{3}^{2}.
\end{eqnarray}
This is the \textit{mini-superspace approximation}.

Furthermore, using the \textit{steepest-descent approximation}, we can regard that the Euclidean path integral can be approximated by the sum-over on-shell solutions (instantons) that satisfy the following equations of motion:
\begin{eqnarray}
\ddot{\phi} &=& - 3 \frac{\dot{\rho}}{\rho} \dot{\phi} \pm V',\\
\ddot{\rho} &=& - \frac{8 \pi}{3} \rho \left( \dot{\phi}^{2} \pm V \right),
\end{eqnarray}
where the upper sign is for the Euclidean time and the lower sign is for the Lorentzian time.
Initially, the equations should satisfy
\begin{eqnarray}
\rho(0) = 0, \;\;\; \dot{\rho}(0) = 1, \;\;\; \dot{\phi}(0) = 0
\end{eqnarray}
for the regularity of equations.

We think that a history initially begins from a Euclidean time and eventually approaches a Lorentzian time. Between two limits, a complex time path connects the initial point and the final point. This path should satisfy the analyticity of complex analysis. Hence, all functions should be complexified in general, although all functions should be realized after a long Lorentzian time. This complexified solutions was named by \textit{fuzzy instantons} \cite{Hartle:2007gi}. Fuzzy instantons should become real-valued functions and this condition is named by \textit{classicality}. Finally, the on-shell Euclidean action becomes
\begin{eqnarray}
S_{\,\mathrm{E}} = 4\pi^{2} \int d \tau \left( \rho^{3} V - \frac{3}{8 \pi} \rho \right)
\end{eqnarray}
and the probability becomes $P \simeq \exp -2 \mathrm{Re} S_{\mathrm{E}}$.

If we only restrict real-valued instantons, then only $\phi=0$ can be a solution, so-called the Hawking-Moss instanton \cite{Hawking:1981fz}. However, if we extend to fuzzy instantons with classicality, then we can find a solution even for $\phi \neq 0$ cases. Then, all the solutions are necessarily dynamic. We need numerical techniques to find the classicalized complexified instantons.

After we find numerical solutions, our purpose is to study the probability distribution. We can try to add the correction term to the action as follows:
\begin{eqnarray}
S_{\mathrm{E}} = -\frac{3}{16 \left( 1 + \mu^{2}\phi^{2}(t_{o})/2 \right)} +C_{\mathrm{E}}(t_{o}) \mu^{2}  \left( \mu\phi(t_{o}) \right)^{2},
\end{eqnarray}
where $t_{\mathrm{o}} \simeq H^{-1} \ln H \rho$ during the Lorentzian time. The first term is the well-known Hawking-Moss instanton term. The second term is the correction term due to the dynamics of the scalar field. Therefore, the second term is the \textit{contribution of fuzzy instantons}. Here, $S_{\mathrm{E}}$ is independent of $t_{\mathrm{o}}$ for a given history, and hence, $C_{\mathrm{E}}(t_{o})$ should have a dependence on $t_{\mathrm{o}}$. Therefore, $t_{\mathrm{o}}$ is not a dynamical time variable; but this corresponds a slice choice in the mini-superspace. We will call this the observing time.

From the numerical observations, the Euclidean action is approximately $S_{\mathrm{E}}=-3/16V(\phi_{\mathrm{T}})$, where $\phi_{\mathrm{T}}$ is the field observed at the \textit{turning point} from the Euclidean time to the Lorentzian time. This was confirmed in \cite{Hwang:2012mf}. Using this, one can discuss by a formal way. Note that we can use the equations of motion in the slow-roll limit:
\begin{eqnarray}
H^{2} &\simeq& \frac{8\pi}{3}V,\\
\dot{\phi} &\simeq& - \frac{V'}{3H}.
\end{eqnarray}
At the observing time $t_{o}$, the potential varies:
\begin{eqnarray}
V(\phi) \simeq V(\phi_{\mathrm{T}}) + V' \dot{\phi} \times t_{o},
\end{eqnarray}
and hence, the Euclidean action can be expanded by
\begin{eqnarray}
S_{\mathrm{E}} &\simeq& -\frac{3}{16 V(\phi_{\mathrm{T}})}\\
&\simeq& -\frac{3}{16 V(\phi)} \left( 1 + \frac{V'}{V} \dot{\phi} \times t_{o} \right)\\
&\simeq& -\frac{3}{16 V(\phi)} + \frac{t_{o}}{16 H} \mu^{4}\phi^{2}.
\end{eqnarray}
Therefore, we obtain $C_{\mathrm{E}} \simeq t_{o}/16H$.

\section{Stochastic approach}

Second, to discuss the quantum field theory in a de Sitter background, we first introduce a de Sitter metric: $ds^{2} = - dt^{2} + a^{2}(t)\textbf{dx}^{2}$. Here, we emphasize that the space topology is flat, while the topology was compact in the Euclidean case (even after the Wick rotation). Therefore, two pictures are completely different at a first glance; while, this difference will give us an important intuition later. Here, $a(t)$ is dimensionless, while $\rho(\tau)$ has a length dimension.

By introducing the conformal time $d\eta = dt/a$ and defining a scalar field as $\varphi \equiv a \phi$,
one can change the equation of motion for $\varphi$:
\begin{equation}
\varphi '' - \nabla^2 \varphi + \left( a^2 m^2 - \frac{a''}{a}   \right) \varphi = 0,
\end{equation}
where a prime is the derivative with respect to $\eta$. In a de Sitter space, $a = e^{Ht}$ and $\eta = - 1/a H$. This equation has the general solution
\begin{equation}
\varphi(\eta, \textbf{x}) = \int \frac{d^3 \textbf{k}}{(2 \pi)^{3/2}}
\left[ a_{\textbf{k}} \varphi_k(\eta) + a^{\dagger}_{-\textbf{k}} \varphi^*_k(\eta) \right] e^{i \textbf{k} \cdot \textbf{x}},
\end{equation}
where $\varphi_k$ satisfies
\begin{equation}\label{eq:varphi_eta}
\varphi''_k +  k^2 \varphi_k + \frac{1}{\eta^{2}} \left( \frac{m^2}{H^2} - 2  \right) \varphi_k = 0.
\end{equation}
Since the solution of Equation~(\ref{eq:varphi_eta}) should have a proper boundary condition in the sub-horizon limit, the solution is expressed as
\begin{equation}
\varphi_k = e^{i \left( \nu + \frac{1}{2} \right)\frac{\pi}{2} }
\sqrt{\frac{\pi}{4k}} \sqrt{-k \eta} H^{(1)}_{\nu}(-k \eta),
\end{equation}
where
\begin{equation}\label{eq:nu}
\nu = \sqrt{\frac{9}{4}-\frac{m^2}{H^2}}.
\end{equation}
Note that if $\nu$ becomes imaginary, then the short wavelength modes cannot be regarded small and hence they cannot be regarded classical. However, if $\nu$ is real, then we can think that short wavelength modes contribute the long wavelength modes (locally homogeneous modes) as a quantum fluctuation.

\begin{figure}[t]
  \resizebox{30pc}{!}{\includegraphics{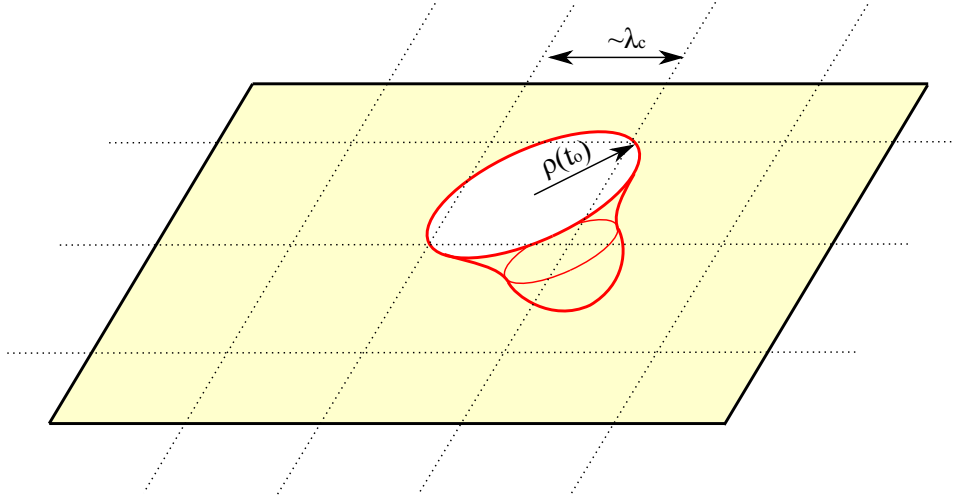}}
\caption{\label{fig:interpretation}Interpretation: the coarse-graining cutoff of stochastic approach corresponds the size of the compact Euclidean instanton at the observing time.}
\end{figure}

Now we calculate the dispersion of $\phi$:
\begin{equation}
\langle \phi^2  \rangle = \frac{1}{a^2} \langle \varphi^2 \rangle
= \frac{1}{a^2} \int \frac{d^3k}{(2\pi)^3} |\varphi_k |^2.
\end{equation}
This integral in itself has a divergence as $k \rightarrow \infty$. Therefore, we need to introduce the ultraviolet cutoff $k_c$. This is consistent with our stochastic approach; we consider only the super-horizon Fourier modes so that $k < k_{c} \sim a H$. Then, the momentum cutoff $k_{c}$ or physical momentum cutoff $q_{c}=k_{c}/a$ corresponds the smoothing length scale (minimal length of relevant waves) $\lambda_{c} \propto 1/q_{c}$ that determines how to coarse-grain the momentum space.

Now, the integration can be expanded as follows:
\begin{eqnarray}\label{eq:stocha}
\langle \phi^2 \rangle \simeq \frac{8}{3 \mu^{2}} \left[ 1 + \frac{\mu^{2}}{4 \pi^{2}} \ln \frac{\tilde{k}_{c}}{aH} + \mathcal{O}\left( \mu^{3} \right) \right],
\end{eqnarray}
where $\zeta \simeq 2.45$ and $\tilde{k}_{c} = k_{c}/\zeta$.
We can represent $\tilde{k}_{c}/H=\exp H(t-\Delta t)$, where $\Delta t$ is a certain constant that we can choose freely. Then
\begin{eqnarray}
\ln \frac{\tilde{k}_{c}}{a H} \simeq - H \Delta t.
\end{eqnarray}

\section{Interpretation}

We can approximate the Euclidean probability as
\begin{eqnarray}
\log P \simeq - 2 S_{\mathrm{E}} \simeq - \frac{\phi^{2}}{2 \langle\phi^{2}\rangle} + ...,
\end{eqnarray}
where this approximation is sufficiently good for the small field limit around the local minimum. Then the dispersion of the field in the Euclidean approach becomes
\begin{eqnarray}
\langle\phi^{2}\rangle \simeq \frac{8}{3\mu^{2}} \left( 1 - \frac{H}{4\pi^{2}} \mu^{2}t_{\mathrm{o}} \right).
\end{eqnarray}
By comparing with the stochastic picture, we conclude that
\begin{eqnarray}
\Delta t = t_{o}
\end{eqnarray}
and
\begin{eqnarray}
\frac{aH}{\tilde{k}_{c}} \simeq \frac{\rho(t_{o})}{\rho(\tau=X)},
\end{eqnarray}
or
\begin{eqnarray}
\frac{1}{\tilde{q}_{c}} \simeq \rho(t_{o}),
\end{eqnarray}
where $\tilde{q}_{c}\equiv\tilde{k}_{c}/a$ is the physical cutoff scale of the momentum space. In the stochastic approach, if a certain quantum fluctuation that has a shorter wave length than $\lambda_{c} \propto 1/\tilde{q}_{c}$, then we can ignore; the universe is homogeneous up to the length scale $\lambda_{c}$. On the other hand, in the Euclidean approach, the instanton is homogeneous up to the size of the universe $\rho(t_{o})$. Now what we can say is that
\begin{eqnarray}
\lambda_{c} \simeq \rho({t_{o}}).
\end{eqnarray}
This is conceptually discussed in Figure~\ref{fig:interpretation}.

This has two important physical meanings: (1) the cutoff is naturally introduced by Euclidean instantons and (2) it is due to \textit{complex-valued} instantons. Therefore, this strongly supports that complex-valued (fuzzy) instantons are indeed physical.

\section{Perspectives}

There were progresses in Euclidean quantum gravity and Euclidean quantum cosmology. We summarize as follows:
\begin{enumerate}
\item Hartle, Hawking and Hertog \cite{Hartle:2007gi} discussed that a classical universe requires inflation. However, the Euclidean probability prefers smaller e-foldings.
\item Hwang, Sahlmann and Yeom \cite{Hwang:2011mp} and Hwang, Lee, Sahlmann and Yeom \cite{Hwang:2011mp2} discussed that complex-valued instantons are useful to explain the initial condition for the stabilization of dilaton or moduli fields.
\item In Hwang, Lee, Sahlmann and Yeom \cite{Hwang:2011mp2}, they commented that the no-boundary wave function perhaps prefers anti de Sitter vacuum.
\item Hwang, Kim, Lee, Sahlmann and Yeom \cite{Hwang:2011mp3} discussed that a large number of fields can help to explain the preference of sufficient e-foldings.
\item Hwang, Lee, Stewart, Yeom and Zoe \cite{Hwang:2012mf} discussed that the complex-valued (fuzzy) instantons are indeed physical by comparing with the stochastic approach. The origin of the cutoff in the stochastic approach can be understood from the Euclidean approach.
\end{enumerate}

However, we need to generalize further points as follows:
\begin{enumerate}
\item It is not entirely clear whether the no-boundary wave function is relevant to explain the realistic inflation model, that may come from particle physics or string theory. This may be connected the viability of inflationary models; also, may be connected to observational consequences.
\item It is not entirely clear how to extend to less symmetric case, for example, spherically symmetric case. This may help us to understand black hole thermodynamics.
\item For large mass cases, there appears a field space such that there is no classicalized universe. What are their meaning, especially in the multiverse? If we think that the multiverse is a statistical realization of the wave function, then we may say something about the unclassical histories of the Euclidean quantum gravity. This may be related to the inhomogeneity of the universe.
\end{enumerate}

The comparison between Euclidean quantum cosmology and loop quantum cosmology can be interesting. In addition, the extension to the spherically symmetric fuzzy instantons can be a challenging issue that may be related to black hole physics. The connection/comparison between the decoherence and the classicality of the instantons - classicalization \textit{without}(?) decoherence - will be useful to understand the fundamental nature of the universe and multiverse.


\begin{theacknowledgments}
The author would like to thank Bum-Hoon Lee, Ewan Stewart, Hanno Sahlmann, Dong-il Hwang, Soo A Kim and Heeseung Zoe for helpful discussions. The author also thanks to Claus Kiefer and Martin Bojowald for helpful comments during this conference. The author is supported by the National Research Foundation of Korea grant funded by the Korean government through the Center for Quantum Spacetime of Sogang University with the grant number 2005-0049409.
\end{theacknowledgments}

\bibliographystyle{aipproc}   

\end{document}